# TED: Towards Discovering Top-$k$ Edge-Diversified Patterns in a Graph Database

Kai Huang [1,2], Haibo Hu [1], Qingqing Ye [1], Kai Tian [3], Bolong Zheng [4], Xiaofang Zhou [2]

[1]Department of Computer Science and Engineering, The Hong Kong University of Science and Technology
[2]Department of Electronic and Information Engineering, Hong Kong Polytechnic University
[3] Tencent; [4] Huazhong University of Science and Technology
ustkhuang|zxf@ust.hk,haibo.hu|qqing.ye@polyu.edu.hk,adamtian@tencent.com,zblchris@gmail.com

## ABSTRACT

With an exponentially growing number of graphs from disparate repositories, there is a strong need to analyze a graph database containing an extensive collection of small- or medium-sized data graphs (*e.g.,* chemical compounds). Although subgraph enumeration and subgraph mining have been proposed to bring insights into a graph database by a set of subgraph structures, they often end up with similar or homogenous topologies, which is undesirable in many graph applications. To address this limitation, we propose the *Top-k Edge-Diversified Patterns Discovery problem* to retrieve a set of subgraphs that cover the maximum number of edges in a database. To efficiently process such query, we present a generic and extensible framework called TED which achieves a guaranteed approximation ratio to the optimal result. Two optimization strategies are further developed to improve the performance. Experimental studies on real-world datasets demonstrate the superiority of TED to traditional techniques.

## 1 INTRODUCTION

The graph database that contains a large collection of small- or medium-sized data graphs has become increasingly prevalent in a variety of domains such as biological networks, drug discovery, and computer vision. Analyzing and mining such a database nurtures many applications, including graph search and classification. While graph analysis varies from application to application, a common phenomenon in these applications is that some subgraph structures (*also known as* graph patterns) play a vital role in characterizing the underlying graph database and building intuitive models for better understanding complex structures. Consequently, subgraph enumeration, which aims to enumerate all subgraphs in a graph database, has been extensively studied in the literature [1–5]. Subgraph enumeration is known to be computationally challenging and memory-consuming since the number of subgraphs in a graph database is exponential to the database size. Therefore, complete enumeration and persistence of all subgraphs in a graph database are infeasible. For instance, more than a million compounds are available from sources such as PubChem [1] and eMolecules [2].

Frequent subgraph mining alleviates this problem by generating only frequent subgraphs instead of all subgraphs [6–11]. Given

---
[1]https://pubchem.ncbi.nlm.nih.gov/
[2]https://www.emolecules.com/

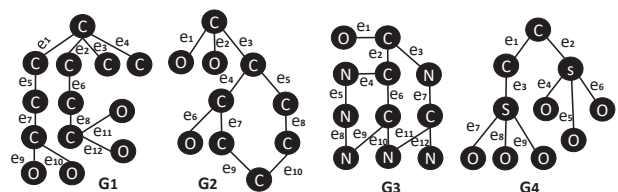
Figure 1: A sample graph database.

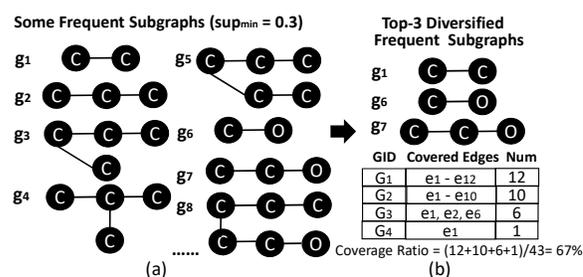
Figure 2: Frequent (resp. diversified frequent) subgraphs.

a minimum support threshold $sup_{min}$ [3], a subgraph $g$ is frequent if the fraction of data graph $G$ containing $g$ (*i.e.,* $g$ is a subgraph of $G$) in a database $D$ is no less than $sup_{min}$. Intuitively, as the threshold decreases, the number of resulting subgraphs increases dramatically. In contrast, the number of subgraphs decreases as the threshold increases, making the resulting subgraphs topologically similar to each other with common substructures and no longer representational. In other words, the results lack diversity whose importance has been advocated in the literature [13–16].

EXAMPLE 1. *To illustrate this, consider a graph database containing $G_1$ to $G_4$ in Figure 1. Let the minimum support $sup_{min}$ = 0.3. The database contains many frequent subgraphs, some of which are shown in Figure 2(a). All these subgraphs except $g_6$ share a common subgraph $g_1$, which is homogenous and thus redundant. A possible way to alleviate this problem is to select a limited number (e.g., k) of diversified ones from all frequent subgraphs such that the maximum number of edges in the database can be covered. For example, if we select the top-3 diversified frequent subgraphs (i.e., $g_1$, $g_6$ and $g_7$) shown in Figure 2(b), they will cover a maximum number of edges (i.e., 67% edges) in all data graphs from $G_1$ to $G_4$. As a comparison, if we just randomly select 3 subgraphs, for example $g_3$, $g_4$ and $g_5$, users cannot obtain any information on graphs $G_3$ and $G_4$ since they contain none of the subgraphs.* ∎

---
[3]$sup_{min} \in [0, 1]$ is a user-specified parameter, which is used to eliminate infrequent subgraph $g$ whose frequency of occurrences in a database $D$ is less than $sup_{min}$.

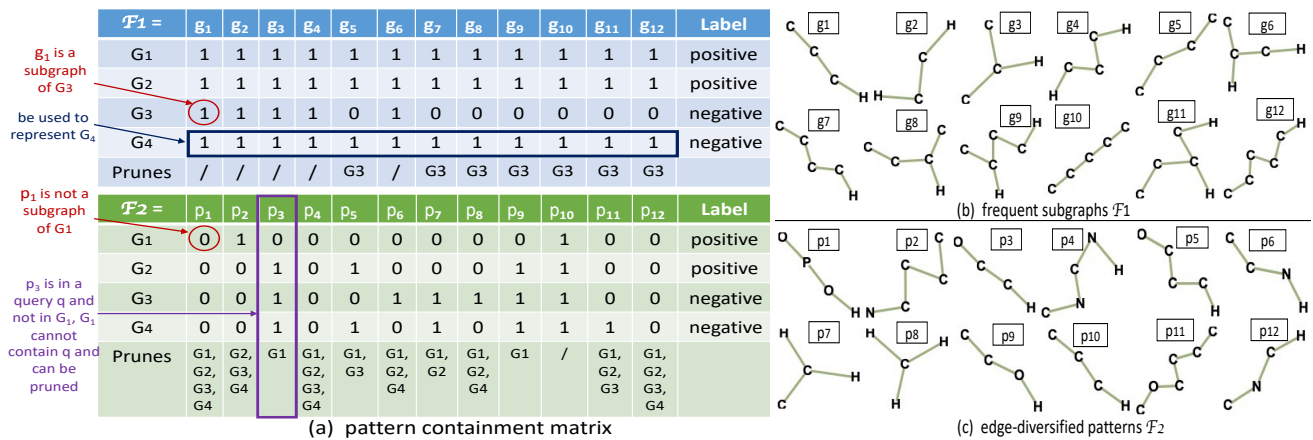

Figure 3: (a) The pattern containment matrix where each entry indicates if a pattern (i.e., $g_i$ or $p_i$) is a subgraph of data graphs, (b) top-$k$ frequent subgraphs $\mathcal{F}_1$, and (c) edge-diversified patterns $\mathcal{F}_2$.

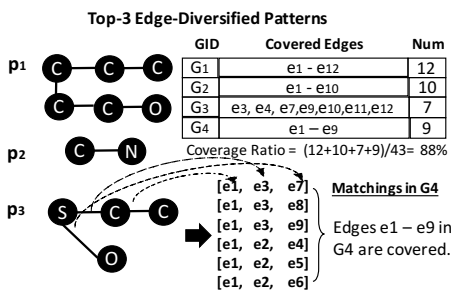

Figure 4: Top-$k$ edge-diversified patterns.

Although these diversified frequent subgraphs $g_1$, $g_6$ and $g_7$ cover all data graphs ($G_1 - G_4$, Figure 1) and 67% of their edges, no edge in $G_4$ (except $e_1$) is covered. Further, $g_7$ does not cover new edges that are not covered by $g_1$ and $g_6$ (see Figure 2(b)). If we replace $g_7$ with an infrequent subgraph that contains nodes "S", "O" and an edge between them, the percentage of covered edges will increase from 67% to 81%. This observation motivates us to retrieve, instead of top-$k$ frequent subgraphs, the top-$k$ diversified subgraphs, also known as edge-diversified patterns, that cover a maximum number of edges in the graph database. Note that edge-diversified patterns can contain both frequent and infrequent subgraphs, as shown in the following example.

Example 2. *Reconsider Example 1. Figure 4 presents top-3 edge-diversified patterns (i.e., $p_1$, $p_2$ and $p_3$) for the sample database in Figure 1. Only $p_1$ is a frequent subgraph while $p_3$ (resp. $p_2$) exists in $G_4$ (resp. $G_3$) only. Although $p_3$ is infrequent, it has 6 matchings in $G_4$ and contributes 9 edges to the total edge coverage, as shown in the figure. As such, the top-3 edge-diversified patterns can cover almost 88% edges in all 4 graphs in the sample database.* ∎

Motivated by the aforementioned examples, in this paper, we study the top-$k$ edge-diversified patterns discovery problem, which finds $k$ connected subgraphs from a graph database such that the maximum number of edges in the graph database can be covered. The main difference between frequent subgraphs and edge-diversified patterns lies in that a frequent subgraph $g$ must be contained by multiple (at least $sup_{min}$) different data graphs even if $g$ has only one occurrence in a data graph $G$, whereas an edge-diversified pattern can be frequent (e.g., $p_1$ in Figure 4) or infrequent (e.g., $p_2$ or $p_3$ in Figure 4), as long as it can reach a wider edge coverage by covering different parts of data graphs with multiple occurrences. In other words, a frequent subgraph is measured independently by its frequency, whereas edge-diversified patterns are jointly measured by the total edge coverage (see Definition 3). Since the latter patterns can supplement the former with infrequent subgraphs covering remaining edges in the graph database, it naturally provides a more comprehensive summary of the database than using frequent subgraphs alone.

Therefore, top-$k$ edge-diversified patterns discovery can be applied in a wide range of applications such as graph indexing and graph classification. Graph indexing and classification have been widely studied in the literature [25, 42] and adopted in different real-world graph databases (e.g., PubChem and eMolecules).

- **Graph Indexing.** Subgraph query, which finds data graphs that contain a query subgraph $q$, is a popular query type. As a subgraph query needs to execute subgraph isomorphism tests (see Definition 1), it is an NP-complete problem [25]. Since it is time-consuming to process a query $q$ over each data graph $G$ in a database, one can build pattern-based graph indices to filter out those data graphs that cannot contain $q$ and only execute subgraph isomorphism tests on the remaining graphs. A graph index is commonly defined as a map from a pattern $p$ to each data graph $G$ that contains this pattern. If a query $q$ contains a particular pattern $p$ while $G$ does not contain $p$, $G$ cannot contain $q$ and can be pruned directly.
- **Graph Classification.** Graph classification, which predicts class labels of data graphs, is an important problem. Despite its increasing importance, many popular feature-based classifiers cannot be applied simply due to the lack of vector representation of graphs. Subgraph patterns can serve as features to vectorize these graphs and embrace feature-based classifiers.

The following examples illustrates the advantage of top-$k$ edge-diversified patterns over top-$k$ frequent patterns for graph indexing and classification. Such advantage also exists in other applications (*i.e.,* visual subgraph query formulation and exploratory subgraph search), which will be elaborated in Section 6.

EXAMPLE 3. *Figures 3(b) and (c) present the top-12 frequent subgraphs ($sup_{min} = 0.3$) $\mathcal{F}_1 = \{g_1, g_2, ..., g_i, ..., g_{12}\}$ and edge-diversified patterns $\mathcal{F}_2 = \{p_1, p_2, ..., p_i, ..., p_{12}\}$ in PUBCHEM database, respectively. $G_1$, $G_2$, $G_3$ and $G_4$ are data graphs where the class label of $G_1$ and $G_2$ is "positive", and that of $G_3$ and $G_4$ is "negative". Figure 3(a) (top, resp. bottom) shows the pattern containment matrix of these data graphs over $\mathcal{F}_1$ (resp. $\mathcal{F}_2$). For example, $G_3$ contains $g_1, g_2, g_3, g_4$ and $g_6$, so the corresponding entries (i.e., 4th row, 2nd-5th and 7th columns) in the matrix are 1.*

*1) Graph Indexing. Consider a common query $q$ [4] that contains all $g_i$ in $\mathcal{F}_1$, $p_3$, $p_5$, and $p_7 - p_{11}$ in $\mathcal{F}_2$. If a graph index $\mathcal{I}$ is built on $\mathcal{F}_1$ (denoted by $\mathcal{I}_{\mathcal{F}_1}$), it can filter out $G_3$. This is because $g_5$ (or $g_7 - g_{12}$) is a subgraph of $q$ but not a subgraph of $G_3$ (Figure 3(a), top). Thus, $q$ is not a subgraph of $G_3$. As such, three subgraph isomorphism tests are needed (i.e., testing $q$ over $G_1$, $G_2$ and $G_4$). Moreover, $\mathcal{I}_{\mathcal{F}_1}$ can reduce at most one subgraph isomorphism test (i.e., $q$ over $G_3$) no matter what the query $q$ is (see the last row of the top table in Figure 3(a)). On the other hand, if the graph index $\mathcal{I}$ is built on $\mathcal{F}_2$ (denoted by $\mathcal{I}_{\mathcal{F}_2}$), it can filter out all these data graphs. This is because $q$ contains $p_3$, $p_5$, and $p_7 - p_{11}$, which jointly filter out $\{G_1, G_2, G_3, G_4\}$. For example, $p_3$ can filter out $G_1$ and $p_5$ can prune $\{G_1, G_3\}$ (see the last row of the bottom table in Figure 3(a)). Therefore, no subgraph isomorphism test is needed.*

*2) Graph Classification. If $\mathcal{F}_1$ is used as vector representations of data graphs, the feature vector of $G_1$ will be $[1, ..., 1]$ since $G_1$ contains all $g_i$ in $\mathcal{F}_1$ (see 2nd row of the top table in Figure 3(a)). If $\mathcal{F}_2$ is used as vector representations of data graphs, the feature vector of $G_1$ will be $[0, 1, 0, ..., 0, 1, 0, 0]$ since $G_1$ contains $p_2$ and $p_{10}$ in $\mathcal{F}_2$ (see 2nd row of the bottom table in Figure 3(a)). Given the $\mathcal{F}_2$-based feature vectors, a simple and yet effective classification rule can be obtained: a graph (e.g., $G_3$ or $G_4$) which contains $\{p_3, p_7, p_9, p_{10}\}$ can be negative. But it is hard to find such a good classification rule with $\mathcal{F}_1$-based feature vectors, since the negative graph $G_4$ has the same representation as that of $G_1$ or $G_2$.*　■

As both frequent (*e.g., $p_3$*) and infrequent patterns (*e.g., $p_8$*) play an important role in graph indexing and graph classification, edge-diversified patterns show their advantage over frequent subgraphs.

However, *top-$k$ edge-diversified patterns discovery* problem introduces non-trivial challenges. First, it is NP-hard. Second, adapting existing subgraph enumeration and frequent subgraph mining techniques to this problem is inadequate since they fall short in handling large databases or guaranteeing patterns' quality. In this paper, we address these challenges and propose a novel solution for top-$k$ edge-diversified patterns discovery problem. We make the following contributions.

- To the best of our knowledge, we are the first to study *top-k edge-diversified patterns discovery problem* and address it with two adapted baseline solutions.
- We further present a generic and extensible framework called TED for this problem which requires limited memory and achieves a guaranteed approximation ratio.

[4]https://pubchem.ncbi.nlm.nih.gov/compound/2244

**Table 1: List of key notations.**

| Notation | Description |
|---|---|
| $g, G, D$ | a subgraph, a graph, a graph database |
| $V(G), E(G)$ | vertex set of graph $G$, edge set of graph $G$ |
| $p, \mathcal{P}$ | a pattern, a pattern set |
| $k$ | number of edge-diversified patterns |
| $sup_{min}$ | minimum support |
| $Cov(G_i, G_j)$ | cover set of $G_i$ over $G_j$ |
| $|Cov(G_i, G_j)|$ | coverage of $G_i$ over $G_j$ |
| $E_{max}$ | maximum number of edges in an edge-diversified pattern |
| $S, S_{fre}$ | a set of subgraphs, a set of frequent subgraphs |

- Two optimization strategies are developed to improve the performance.
- By using real-world data graph repositories, extensive experimental evaluations are provided to show the superiority of our methods over two baseline solutions.

The rest of this paper is organized as follows. In Section 2, we provide some preliminaries and the problem statement. In Section 3, two baseline solutions are proposed for top-$k$ edge-diversified patterns discovery problem. In Section 4, our proposed novel framework TED is presented, followed by two optimization strategies in Section 5. In Section 6, we employ the demonstration system called VINCENT to illustrate the application potentials of edge-diversified patterns. Section 7 shows extensive experimental results. Related work are in Section 8 and conclusions are made in Section 9.

## 2 PRELIMINARIES

Table 1 lists the notations and acronyms used in this paper.

### 2.1 Key Concepts

A simple graph $G$ is represented as $G = (V, E)$ where $V$ is a set of vertices and $E \subseteq V \times V$ is a set of edges. For ease of presentation, we assume $G$ is an undirected connected graph whose vertex $v \in V$ is labeled with $l(v)$ and edge $e \in E$ is labeled with $l(e)$ [5]. In this paper, we focus on a graph database containing a large collection of graphs (denoted as $D$), each of which is with either dozens or hundreds of nodes (up to 801, see Table 2, Section 7.1). Therefore, this paper follows related works [16, 18] to call them small- or medium-sized graphs. A unique *index* (*i.e.,* ID) is assigned to each graph in $D$. We denote a graph with index $i$ as $G_i \in D$.

DEFINITION 1 (SUBGRAPH ISOMORPHISM). *Given two graphs $G_1$ and $G_2$, a subgraph isomorphism is an injection $f: V(G_1) \rightarrow V(G_2)$ such that 1) $\forall v \in V(G_1), l(v) = l'(f(v))$ and 2) $\forall (u, v) \in E(G_1), (f(u), f(v)) \in E(G_2)$ and $l(u, v) = l'(f(u), f(v))$ where $l$ and $l'$ are the labeling functions of graph $G_1$ and $G_2$, respectively.*

$G_1$ is *subgraph isomorphic to* $G_2$ if there exists at least one subgraph isomorphism $f$ from $G_1$ to $G_2$. We also say that $G_2$ is *covered* by $G_1$ or that $G_2$ *contains* $G_1$ (denoted by $G_1 \subseteq G_2$).

DEFINITION 2 (COVER SET AND COVERAGE). *Given two graphs $G_1 = (V_1, E_1)$ and $G_2 = (V_2, E_2)$, if $G_1$ is subgraph isomorphic to $G_2$ and the matchings are $\mathcal{F}$, the cover set of $G_1$ over $G_2$ is $Cov(G_1, G_2) = \cup_{f \in \mathcal{F}} (f(u), f(v))$ and the coverage is $|Cov(G_1, G_2)|$.*

[5]For the graph with labeled vertex and unlabeled edge, each edge $e \in E$ is labeled with the concatenation of labels (*i.e.,* a new label joining two labels together) of its two end vertices (*i.e.,* $l(e) = l(u).l(v)$).

## 2.2 Problem Statement

DEFINITION 3 (TOP-k EDGE-DIVERSIFIED PATTERNS DISCOVERY). *Given a graph database $D = \{G_1, G_2, \cdots, G_j, \cdots, G_n\}$ and an integer $k$, top-$k$ edge-diversified patterns discovery is to find $k$ connected subgraphs $\mathcal{P} = \{p_1, p_2, \ldots, p_i, \ldots p_k\}$ from $D$ such that the total coverage of $\mathcal{P}$ over $D$ (denoted by $|Cov(\mathcal{P}, D)|$), i.e., $|\cup_i \cup_j Cov(p_i, G_j)|$, is maximized, where $Cov(p_i, G_j)$ is the cover set of $p_i$ over $G_j$, and the number of edges of $p_i$ (i.e., $|E(p_i)|$) is no more than a given size threshold $E_{max}$.*

**Remark.** It seems that when $E_{max} = 1$, this problem becomes the (minimum) edge cover problem, which finds a set of edges $\mathbb{C}$ with the smallest possible size such that each vertex in a graph is incident with at least one edge in $\mathbb{C}$. However, given a graph $G$ (i.e., $D = \{G\}$), even if $E_{max} = 1$, they are not the same, as edge-diversified patterns discovery is to find a limited number of edges (i.e., $k$) such that their matchings in $G$ cover the maximum number of edges, while the (minimum) edge cover problem finds an edge set that covers *all vertices* in $G$. For example, let $k = 1$ and $D = \{G_1\}$ in Figure 1, top-$k$ edge-diversified patterns are the edge between two "C", but (minimum) edge cover of $G_1$ is $\{e_3 - e_6, e_9 - e_{12}\}$. In addition, note that $|E(p_i)|$ is not allowed to be larger than a given threshold $E_{max}$. The reasons are two-fold. First, if some graphs in $D$ are apparently larger than other graphs, there is a likelihood that selecting those graphs as the top-$k$ edge-diversified patterns is already an optimal solution. Thus, we don't have to discuss the problem here. Second, in most applications, such as visual subgraph query formulation, patterns should not be too large [16], since pattern budget (e.g., minimum size, maximum size and the number of patterns) is pre-defined for pattern mining.

THEOREM 1. *The top-k edge-diversified patterns discovery problem is NP-hard.*

PROOF. Let $k = 1$, the original top-$k$ edge-diversified patterns discovery problem (TED Problem) becomes to find a single edge-diversified pattern, i.e., a graph $g$ that covers the maximum number of edges. The reformulated problem (denoted by Simplified TED Problem), i.e., TED Problem with $k = 1$, can be reduced from the maximum coverage problem [22], which is a classical NP-hard optimization problem. □

## 3 BASELINE SOLUTIONS

In this section, we present two baseline solutions for this problem. The first one, named as $\text{ALL}_g$, is to enumerate all subgraphs $S$ from the database $D$, and then conduct a greedy search. Given a minimum support threshold $sup_{min}$, the second solution named as $\text{FSG}_g$ first generates all frequent subgraphs $S_{fre}$ whose support are no less than $sup_{min}$, instead of all subgraphs. It then adopts the same greedy strategy as $\text{ALL}_g$ to find top-$k$ edge-diversified patterns.

### 3.1 Baseline Solution $\text{ALL}_g$

By adopting greedy search, we come up with the first baseline solution $\text{ALL}_g$, whose pseudo-code is shown in Algorithm 1. Given a graph database $D = \{G_1, G_2, \ldots G_n\}$ and an integer $k$, it first enumerates all subgraphs $S = \{s_1, s_2, \ldots\}$ from $D$ such that $|E(s_i)| \leq E_{max}$ (Line 1), by using an existing subgraph enumeration method [1–5].

---

**Algorithm 1** Baseline solution $\text{ALL}_g$

**Input:** graph database $D = \{G_1, G_2, \ldots G_n\}$, integer $k$, and $E_{max}$
**Output:** Near-optimal top-$k$ edge-diversified patterns
1: $S \leftarrow \text{ENUMALLSUB}(D, E_{max})$
2: $\mathcal{P} \leftarrow \text{MAXCOVER}(S, k)$                     ▷ call procedure MAXCOVER
3: **return** $\mathcal{P}$
4: **procedure** MAXCOVER($S, k$)
5:     $\mathcal{P} \leftarrow \phi$
6:     **for** iter = 1 to k **do**
7:         $p \leftarrow argmax_{p' \in S}|Cov(p', D) \setminus Cov(\mathcal{P}, D)|$
8:         $\mathcal{P} \leftarrow \mathcal{P} \cup \{p\}, S \leftarrow S \setminus p$
9:     **return** $\mathcal{P}$

---

**Algorithm 2** Baseline solution $\text{FSG}_g$

**Input:** graph database $D = \{G_1, G_2, \ldots G_n\}$, integer $k$, and $E_{max}$
**Output:** Heuristic top-k edge-diversified patterns
1: $S_{fre} \leftarrow \text{ENUMFRESUB}(D, E_{max})$
2: $\mathcal{P} \leftarrow \text{MAXCOVER}(S_{fre}, k)$               ▷ MAXCOVER in Algorithm 1
3: **return** $\mathcal{P}$

---

Then, MAXCOVER is invoked to generate top-$k$ edge-diversified patterns (Line 2). MAXCOVER follows a iterative procedure (Lines 6-8). In each iteration, it greedily selects a pattern $p$ from $S$ such that the cover set of $p$ contains a maximum number of uncovered edges, i.e., $|Cov(p', D) \setminus Cov(\mathcal{P}, D)|$ (Line 7). Once the resulting pattern $p$ is selected, it will be removed from $S$ to pattern set $\mathcal{P}$ (Line 8).

Observe that MAXCOVER follows the same greedy strategy used for solving the max $k$-cover problem [22]. Therefore, it can achieve an approximation ratio of $1 - \exp(-1)$.

LEMMA 1. *Worst case time and space complexities of $\text{ALL}_g$ are $O(|D|2^{max(V(G))^2} + k|S||D|max(V(G))^{E_{max}})$ and $O(max(E(G))|D| + E_{max}|S|)$, respectively, where $max(V(G))$ (resp. $max(E(G))$) is maximum number of vertices (resp. edges) in graph $G \in D$, and $|D|$ (resp. $|S|$) is number of graphs in $D$ (resp. $S$).*

### 3.2 Baseline Solution $\text{FSG}_g$

Note that $\text{ALL}_g$ requires searching all subgraphs which may incur large computational overload. To address this, we further propose the second baseline solution $\text{FSG}_g$, whose procedure is shown in Algorithm 2. Instead of enumerating all subgraphs from graph database $D$, $\text{FSG}_g$ first adopts frequent subgraph mining methods [6–11] to generate frequent subgraphs $S_{fre}$ (Line 1). Then MAXCOVER in Algorithm 1 is invoked to find a pattern set on $S_{fre}$ (Line 2).

LEMMA 2. *Worst case time and space complexities of $\text{FSG}_g$ are $O(|D|2^{max(V(G))^2} + k|S_{fre}||D|max(V(G))^{E_{max}})$ and $O(max(E(G))|D| + E_{max}|S_{fre}|)$, respectively, where $max(V(G))$ (resp. $max(E(G))$) is maximum number of vertices (resp. edges) in graph $G \in D$, and $|D|$ (resp. $|S_{fre}|$) is number of graphs in $D$ (resp. $S_{fre}$).*

### 3.3 Limitations of Baseline Solutions

Observe that $\text{ALL}_g$ provides an approximation ratio of $1 - \exp(-1)$ for final patterns, but it is computationally challenging. Compared to $\text{ALL}_g$, $\text{FSG}_g$ significantly reduces the computational cost by selecting frequent subgraphs, based on which the final patterns are derived. However, the pattern quality is not guaranteed since no approximation bound can be provided as only frequent subgraphs are considered. Moreover, they suffer from the following problems:

(1) *Excessive memory consumption.* Both $\text{ALL}_g$ and $\text{FSG}_g$ need to store all subgraphs or a subset of all subgraphs in memory. However, the number of subgraphs in a graph database is exponential in database size. As a graph database continues to grow rapidly in size, storing those subgraphs in memory will lead to excessive memory consumption.

(2) *A mass of unnecessary computations.* Since the (frequent) subgraph enumeration and MaxCover procedure are performed sequentially, all (frequent) subgraphs, no matter whether they can improve the total coverage of final patterns $\mathcal{P}$, are computed and stored before invoking MaxCover procedure. Therefore, there are a mass of unnecessary computations.

## 4 TED: A NOVEL FRAMEWORK

To address aforementioned limitations, we propose a novel framework called TED (*i.e.*, Top-$k$ Edge-Diversified patterns discovery). The main idea is to integrate the search process for top-$k$ patterns into subgraph enumeration. Specifically, it maintains only $k$ candidate patterns $\mathcal{P}$ in memory. These candidates are supposed to potentially maximize the total coverage. When a new subgraph $g$ is enumerated and considered as a *promising candidate* (see Sec. 4.1), it will be added to $\mathcal{P}$ by swapping out one of the patterns. Compare to the above two baseline solutions, TED has the following advantages:

(i) *Limited memory consumption.* TED always maintains only $k$ patterns in memory to avoid excessive memory consumption for storing all subgraphs. As can be seen in Theorem 2, the space complexity of TED is $O(max(E(G))|D|)$ which is far less than that of $\text{ALL}_g$ ($O(max(E(G))|D| + E_{max}|S|)$, Lemma 1) and $\text{FSG}_g$ ($O(max(E(G))|D| + E_{max}|S_{fre}|)$, Lemma 2).

(ii) *Guaranteed pattern quality.* TED can achieve an approximation bound of 1/4 and better performance in experimental study. To this end, it provides a swapping criteria such that each newly generated subgraph $g$ is deliberately swapped with one existing pattern in $\mathcal{P}$.

(iii) *Effective pruning strategies.* TED is able to prune subgraph search space by integrating the search process for top-$k$ patterns into subgraph enumeration.

In what follows, we begin with a basic TED algorithm (denoted by TED_BASE) in this section, which can achieve both (i) and (ii). Then, two optimization strategies are developed to further realize (iii) (Section 5).

### 4.1 The Basic TED Algorithm

The basic TED algorithm (TED_BASE) generates top-k edge-diversified patterns by alternately performing subgraph enumeration and a search process for top-$k$ patterns. For subgraph enumeration, it adopts a depth-first search (DFS) strategy to traverse the search space. For example, $g_{1,1}, g_{2,1}, g_{3,1}$, and $g_{4,1}$ in Figure 5 are traversed sequentially. Although Apriori-like approaches [6, 7] that adopt breadth-first search (BFS) strategy have been widely studied, they require generating a lot of duplicated candidates and testing subgraph isomorphism. In contrast, depth-first search (DFS) combines the pattern generating and isomorphism checking into one process to address this problem. For top-$k$ pattern search on the enumerated subgraphs, a swapping-based strategy is adopted for maintaining patterns with limited memory consumption.

TED_BASE is outlined in Algorithm 3. Given a graph database $D = \{G_1, G_2, ...G_n\}$ and an integer $k$, it first enumerates all 1-sized

---

**Algorithm 3** A basic TED algorithm (TED_BASE)

**Input:** graph database $D = \{G_1, G_2, ...G_n\}$, integer $k$, and $E_{max}$
**Output:** Near-optimal top-k edge-diversified patterns
1: $\mathcal{P} \leftarrow \phi$
2: $S_p \leftarrow \text{EnumSub}(D, |E| = 1)$
3: **for** each $g \in S_p$ **do**
4: $\quad S_p \leftarrow S_p \setminus g$
5: $\quad \mathcal{P} \leftarrow \text{PatternMaintain}(\mathcal{P}, g, D)$
6: $\quad \mathcal{P}_g \leftarrow \text{RightMostExtend}(g, D, E_{max})$
7: $\quad S_p \leftarrow \mathcal{P} \cup \mathcal{P}_g$
8: **procedure** PatternMaintain($\mathcal{P}, g, D$)
9: $\quad$ **if** $|\mathcal{P}| < k$ **then**
10: $\quad\quad \mathcal{P} \leftarrow \mathcal{P} \cup \{g\}$
11: $\quad\quad$ **return** $\mathcal{P}$
12: $\quad \text{Score}_L, p_t \leftarrow \text{MIN}(\text{genLossScore}(\mathcal{P}))$
13: $\quad \text{Score}_B \leftarrow \text{genBenefitScore}(g)$
14: $\quad$ **if** $\text{Score}_B > (1 + \alpha)\text{Score}_L + \frac{(1-\alpha)|Cov(\mathcal{P},D)|}{k}$ **then**
15: $\quad\quad \mathcal{P} \leftarrow \mathcal{P} \setminus p_t \cup \{g\}$
16: $\quad$ **return** $\mathcal{P}$

---

subgraphs (*i.e.*, edges) and appends them into the set $S_p$ (Line 2). Then, an iterative process (Lines 3-7) is performed to generate final patterns $\mathcal{P}$ by taking $S_p$ and $D$ as inputs. Specifically, each subgraph $g \in S_p$ is first considered and removed from $S_p$ (Lines 3-4). Then, the procedure PatternMaintain is performed to update top-$k$ edge-diversified patterns $\mathcal{P}$ with newly enumerated subgraph $g$ (Line 5). After that, the right-most extension method (Procedure RightMostExtend) extends each subgraph $g \in S_p$ with one more edge (Line 6) so that its supergraphs will be considered in the next iteration. The process repeats until $S_p$ is empty. More details about PatternMaintain and RightMostExtend are discussed below.

*4.1.1 Pattern Maintenance (PatternMaintain).* As discussed above (Section 3), the max $k$-cover problem is a subproblem of top-$k$ edge-diversified patterns discovery. However, greedy-search based solutions typically find entire subgraphs and store them in memory and hence cannot be effectively exploited for large databases. To address this, PatternMaintain maintains only $k$ patterns in memory with a swapping-based method motivated by existing maximum coverage solver in the context of streaming scenario [23–25]. We first introduce the concepts of *loss score* and *benefit score* below to facilitate exposition.

**Definition 4 (Loss score).** Given a pattern set $\mathcal{P}$ and a database $D$, the **loss score** of a pattern $p \in \mathcal{P}$ is the decrease of total coverage caused by removing $p$ from $\mathcal{P}$, i.e.,

$$\text{Score}_L(p, \mathcal{P}, D) = |\cup_{p \in \mathcal{P}} Cov(p, D) \setminus \cup_{p' \in \mathcal{P} \setminus p} Cov(p', D)|.$$

**Definition 5 (Benefit score).** Given a pattern set $\mathcal{P}$ and a database $D$, the **benefit score** of a pattern $g \notin \mathcal{P}$ is the increase of total coverage caused by adding $g$ to $\mathcal{P}$, i.e.,

$$\text{Score}_B(g, \mathcal{P}, D) = |\cup_{p' \in \mathcal{P} \cup \{g\}} Cov(p', D) \setminus \cup_{p \in \mathcal{P}} Cov(p, D)|.$$

PatternMaintain first greedily selects $k$ patterns into the pattern set $\mathcal{P}$ (Lines 9-11, Algorithm 3). When a new subgraph $g$ is generated, a swapping-based process is developed to determine if $g$ should be swapped into $\mathcal{P}$ (Lines 12-15). Specifically, it first calculates and ranks loss scores for each pattern $p \in \mathcal{P}$, and then records the pattern $p_t$ and its pattern score $\text{Score}_L$ such that $p_t$ has a minimum loss score (Line 12). Meanwhile, the benefit score $\text{Score}_B$ of $g$ is also recorded (Line 13). The subgraph $g$ is considered as a **promising candidate** and swapped into $\mathcal{P}$ (Lines 14-15) if the

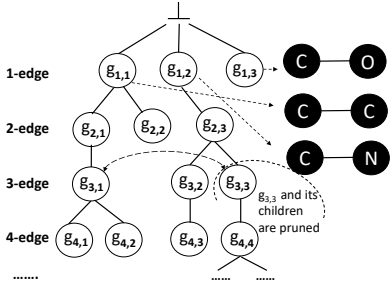

**Figure 5: The DFS search space.**

following *swapping criteria* is satisfied:

$$\text{Score}_B > (1 + \alpha)\text{Score}_L + (1 - \alpha)|Cov(\mathcal{P}, D)|/k \quad (1)$$

where $\alpha \in [0, 1]$ is a *swapping threshold* to balance $\text{Score}_L$ and the average coverage of the patterns in $\mathcal{P}$ (i.e., $|Cov(\mathcal{P}, D)|/k$). Note that there are three variants of the swapping criteria, namely $Swap_1$ [23], $Swap_2$ [24], and $Swap_\alpha$ [25] where $\alpha = 1, \alpha = 0$, and $\alpha \in (0, 1)$, respectively. By default, we set $\alpha = 1$ and compare it with other variants in Section 7. The pattern $p_t$ is swapped out once $g$ is swapped in. The pattern set $\mathcal{P}$ is hence updated (Line 15).

*4.1.2 Right-Most Extension (RightMostExtend).* Ted_base adopts a depth-first search (DFS) strategy to enumerate all possible subgraphs by iteratively performing right-most extension [9] (i.e., RightMostExtend). In particular, given a graph $g$, we can perform DFS on it and generate the corresponding DFS trees. The graph $g$ subscripted with a DFS tree $T$ is denoted by $g_T$ and hence $T$ is named a DFS subscripting of $g$. Given $g_T$, *root vertex* is the first visited vertex (i.e., $V_0$) and *right-most vertex* is the last visited vertex (i.e., $V_{|V(g)|-1}$). The *right-most path* is then defined as the straight path from root vertex to right-most vertex. Moreover, the *forward edge* contains all edges in the DFS tree $T$ (denoted by $E_T^f$). The *backward edge* consists of all other edges (denoted by $E_T^b$). It is obvious that $(V_i, V_j) \in E_T^f$ if $i < j$, and $(V_i, V_j) \in E_T^b$ otherwise.

**Definition 6 (Right-most extension).** Given an $m$-edge graph $g$ and a DFS tree $T$, right-most extension is to extend $g$ to an $(m + 1)$-edge graph $g'$ with an edge $e$ such that at least one of the following rules is satisfied: 1) *forward extension*. $e$ is extended from vertices on the right-most path such that an additional vertex is introduced and 2) *backward extension*. $e$ is extended from right-most vertex to other vertices on the right-most path.

Figure 5 outlines the DFS search space, where each node represents an $m$-edge graph (denoted by $g_{m,i}$, i.e., $i$-th graph with $m$ edges). Each link between two nodes represents a possible right-most extension. For example, $g_{1,1}, g_{2,1}, g_{3,1}$, and $g_{4,1}$ in Figure 5 are traversed sequentially. $g_{4,1}$ is a right-most extension of $g_{3,1}$.

### 4.2 Fast Pattern Maintenance

A naive method for computing loss score (Line 12, Algorithm 3) and benefit score (Line 13, Algorithm 3) is to directly calculate cover set $Cov(\cdot)$ according to their definitions, which is obviously inefficient. In this section, we propose the FastMaintain algorithm to facilitate fast pattern maintenance. To begin with, the Private-Edge-Set Index (denoted by PES-Index) is developed to accelerate loss score and benefit score computation. PES-Index consists of five components:

- $|Cov(\mathcal{P})|$: total coverage of $\mathcal{P}$ over $D$ (i.e., $|Cov(\mathcal{P}, D)|$).
- $|pCov(p)|$: *private coverage* of $p$. $pCov(p)$ is the set of edges in $D$, which is in the cover set of $p$ but not in the cover set of $p' \in \mathcal{P} \setminus p$, i.e., $pCov(p) = Cov(p, D) \setminus Cov(\mathcal{P} \setminus p, D)$.
- $rCov(e)$: *reverse cover set* of an edge $e$. It refers to a subset of patterns that contain $e$ in the cover set. That is, $rCov(e) = \{p|p \in \mathcal{P}, e \in Cov(p, D)\}$.
- $rCnt(i)$: *reverse counting set* of number of edges. It refers to a set of patterns such that each pattern $p$'s private coverage is $i$. That is, $rCnt(i) = \{p|p \in \mathcal{P}, |pCov(p)| = i\}$ where $i \in [0, \sum_{G \in D} |E(G)|]$.
- $p_{min}$: the pattern $p \in \mathcal{P}$ with minimum private coverage $|pCov(p)|$.

Moreover, four operations, Insert, Delete, Update and Select, are developed on PES-Index. Specifically, Insert operation is to insert a newly enumerated subgraph into current pattern set and update PES-Index; Delete aims to remove a pattern from current pattern set and update PES-Index; Update is a combination operation that sequentially calls Delete and Insert operations; Select supports fast computations for the loss score and benefit score.

Given a graph database $D$, a pattern set $\mathcal{P}$ and each enumerated graph $g$, if the number of existing patterns is less than $k$, Insert operation is performed to update $\mathcal{P}$, and $g$ is directly inserted into $\mathcal{P}$. For each edge $e$ covered by $g$ (i.e., $e \in Cov(g, D)$), its reverse cover set is updated by adding $g$. If its reverse cover set only contains $g$ (i.e., $|rCov(e)|==1$), it indicts that $e$ is only covered by $g$. Hence, both g's private coverage $|pCov(g)|$ and the total coverage $Cov(\mathcal{P})$ increase by 1. If its reverse cover set contains another pattern $p$, the private coverage of $p$ (i.e., $|pCov(p)|$) should be decreased by 1 since $e$ is also covered by $g$. Once $|pCov(p)|$ is updated, the reverse counting set $rCnt(i)$ is updated by moving $p$ from $rCnt(|pCov(p)| + 1)$ to $rCnt(|pCov(p)|)$. The process repeats until all $e \in Cov(g, D)$ are considered. Then, $rCnt(|pCov(g)|)$ is also updated. Based on all computed $rCnt(i)$, we can directly figure out which pattern has the minimum loss score by selecting one pattern $p_{min}$ in $rCnt(i) \neq \phi$ such that $i$ is minimum. With the constructed PES-Index, the minimum loss score (i.e., $\text{Score}_L$) and the corresponding pattern (i.e., $p_t$) can be easily obtained ($\text{Score}_L = |pCov(p_{min})|$ and $p_t = p_{min}$). The benefit score $\text{Score}_B$ is calculated by counting the cases where the reverse cover set $|rCov(e)|$ of $e \in Cov(g, D)$ is 0, as $|rCov(e)| = 0$ indicates that $e$ is not covered by existing patterns.

If the swapping criterion is satisfied, the Update operation is performed by sequentially calling Delete operation for $p_t$ and Insert operation for $g$. Firstly, $p_t$ is directly removed from $\mathcal{P}$ and $rCnt(|pCov(p_t)|)$. For each edge $e \in Cov(p_t, D)$, the corresponding $rCov(e)$ should be maintained by removing $p_t$, and the total coverage should be decreased by 1 if $|rCov(e)|$ is 0. In addition, if the $|rCov(e)|$ of $e$ contains only one pattern $p$, the $|pCov(p)|$ should be increased by 1, and the reverse counting set $rCnt(i)$ should be updated by moving $p$ from $rCnt(|pCov(p)|-1)$ to $rCnt(|pCov(p)|)$.

**Remark.** Note that PES-Index is motivated by the existing PNP-Index [46], which was designed for diversified top-k clique search. The main difference lies in that PNP-Index is to build the relationship between the enumerated clique and the vertices in a single large data graph. Each clique is exactly one matching in the data graph and only the vertices in the matching are indexed. In contrast,

PES-Index aims to index the cover set (*i.e.*, a set of edges) of a particular edge-diversified pattern over a graph database containing a set of graphs. Each pattern here has multiple matchings in a graph.

EXAMPLE 4. *Let $\alpha = 1$, $k = 3$, and the current pattern set $\mathcal{P} = \{g_1, p_1, p_3\}$ where $g_1$ and $p_1$ (resp. $p_3$) are shown in Figure 2(a) and Figure 4, respectively. $|pCov(g_1)|$, $|pCov(p_1)|$ and $|pCov(p_3)|$ are 2, 10, and 8, respectively, where $pCov(g_1) = \{G_3 : e_2, e_6\}$ (i.e., $e_2$ and $e_6$ of $G_3$). In addition, $|Cov(\mathcal{P})| = 33$ and $rCnt(2) = \{g_1\} \neq \phi$. Hence, $p_t = g_1$ and the minimum loss score $SCORE_L = |pCov(g_1)| = 2$. When the graph $p_2$ (Figure 4) is newly enumerated, its benefit score $SCORE_B = 7$, as $Cov(p_2, D) = \{G_3 : e_3, e_4, e_7, e_9 - e_{12}\}$ and $|\{e|e \in Cov(p_2, D) \text{ and } |rCov(e)| == 0\}| = 7$. Since $SCORE_B > (1+\alpha)SCORE_L = 4$, $g_1$ should be removed from $\mathcal{P}$ and then $|Cov(\mathcal{P})| = 31$. In addition, $p_2$ should be added into $\mathcal{P}$, i.e., $\mathcal{P} = \{p_1, p_2, p_3\}$. The total coverage is $|Cov(\mathcal{P})| = 38$ as shown in Figure 4.* ∎

THEOREM 2. *Worst case time and space complexities of TED_BASE are $O(|D|2^{max(V(G))^2})$ and $O(max(E(G))|D|)$, respectively, where $max(V(G))$ (resp. $max(E(G))$) is maximum number of vertices (resp. edges) in graph $G \in D$, and $|D|$ is number of graphs in $D$.*

PROOF. The main time cost is spent on ENUMSUB($D, |E| = 1$) (Line 2, Algorithm 3), PATTERNMAINTAIN($\mathcal{P}, g, D$) (Line 5) and RIGHTMOSTEXTEND($g, D, E_{max}$) (Line 6). ENUMSUB($D, |E| = 1$) is to enumerate all edges and thus takes $O(|D|max(E(G)))$ time. PATTERNMAINTAIN($\mathcal{P}, g, D$) requires computations for loss score and benefit score, which take $O(|D|max(E(G)))$ time. In the worst case, total time cost of RIGHTMOSTEXTEND($g, D, E_{max}$) is taken on enumerating all subgraph, which takes $O(|D|2^{max(V(G))^2})$ time. Overall, the worst case time complexity is $O(|D|2^{max(V(G))^2})$, since $|D|2^{max(V(G))^2} \gg |D|max(E(G))$. The space cost is mainly spent on storing the database $D$, pattern set $\mathcal{P}$ and PES-Index. It is obviously that the former two take $O(max(E(G))|D|)$ and $O(kE_{max})$ space, respectively. For PES-Index, the main space is spent on storing reverse cover set $rCov(e)$, which consumes $O(max(E(G))|D|)$ space. Therefore, the worst case space complexity is $O(max(E(G))|D|)$ since $max(E(G))|D| \gg kE_{max}$. □

## 5 OPTIMIZATIONS

Recall that TED_BASE integrates the search process for top-$k$ patterns into subgraph enumeration, which guarantees the pattern quality with limited memory consumption. However, it does not explore the potential of reducing the search space and unnecessary computations to boost the overall effectiveness. Therefore, in this section, we further propose two optimization strategies, namely, *Promising Right-Most Extension* in Section 5.1, and *Initial Pattern Selection* in Section 5.2. In short, *Promising Right-Most Extension* prunes unpromising patterns based on the covering relationship of a graph and its descendants. *Initial Pattern Selection* targets on promoting the quality of initial patterns so that better pruning power is introduced in the early stage.

### 5.1 Promising Right-Most Extension (TED_PRM)

Recall that in TED_BASE, graphs are enumerated with right-most extension. As shown in Figure 6(a), an $m$-edge graph $g$ is extended to an $(m+1)$-edge graph $g'$. Then, $g'$ is extended to an $(m+2)$-edge graph $g''$. It is obvious they discard the relationship (see Observation I) between a subgraph and its descendants.

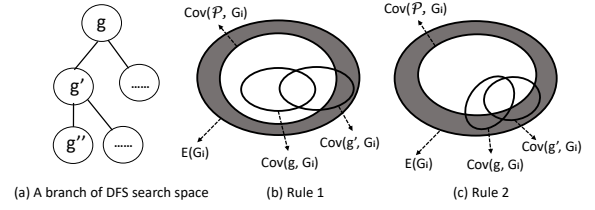

Figure 6: Illustration for PRM.

OBSERVATION I. *The benefit score of a subgraph $g'$ over graph $G_i$ is at most the number of uncovered edges in $G_i$, and the edges in a graph $G_i$ covered by $g$ but not covered by $g'$ will not be covered by $g''$ (i.e., a descendant of $g'$).*

Motivated by this observation, *Promising Right-Most Extension* (PRM) is developed by taking a prudent strategy to extend $g$ to $g'$ (and descendants of $g'$, *e.g.*, $g''$). Specifically, PRM extends $g$ to $g'$ if and only if one of the following PRM rules is satisfied.

DEFINITION 7 (**PRM RULES**). *1) Rule 1. If $g \in \mathcal{P}$, $g'$ extends from $g$ with one edge if $|\cup_{i \in \mathbb{I}} (E(G_i) \setminus Cov(\mathcal{P}, G_i))| \geq (1+\alpha)SCORE_L + (1-\alpha)|Cov(\mathcal{P}, D)|/k$, where $\mathbb{I}$ is the id set of graphs containing $g$. 2) Rule 2. If $g \notin \mathcal{P}$, $g'$ extends from $g$ with one edge if $|\cup_{i \in \mathbb{I}} (E(G_i) \setminus (Cov(\mathcal{P}, G_i) \cup (Cov(g, G_i) \setminus Cov(g', G_i))))| \geq (1+\alpha)SCORE_L + (1-\alpha)|Cov(\mathcal{P}, D)|/k$.*

THEOREM 3. *Right-most extension with PRM rules has no effect on the quality (i.e., coverage) of final patterns.*

PROOF. Let the current patterns be $\mathcal{P}$ and final patterns $\mathcal{P}_{final}$, according to the swapping criteria (Equ. (1), Section 4), only the promising candidate $g$ could be kept in $\mathcal{P}_{final}$. Therefore, given a graph $g$, we prove that PRM rules have no effect on the quality (*i.e.*, coverage) of final patterns by showing that no promising candidate is filtered out. □

Owing to the pruning power of PRM rules, we propose an improved TED algorithm called TED_PRM by introducing PRM into TED_BASE. Instead of executing lines 6 and 7 in Algorithm 3 sequentially, given an $m$-edge graph $g$ and its all potential $(m + 1)$-edge supergraphs $\mathcal{P}_g$, for each $g' \in \mathcal{P}_g$, if $g'$ does not satisfy PRM rules, $g'$ will be removed from $\mathcal{P}_g$ for further processing.

### 5.2 Initial Pattern Selection (TED_IPS)

As discussed in Section 4.1, the initial $k$ patterns are selected by traversing the search space in *DFS* manner (Line 6, Algorithm 3). For example, suppose the search space is shown in Figure 5 and $k = 3$, the initial $k$ patterns are $g_{1,1}, g_{2,1}$, and $g_{3,1}$. Observe that all of them contain the same substructure (*i.e.*, $g_{1,1}$), they may cover the same edges in $D$. Therefore, the minimum loss score $SCORE_L$ may get small, so that the swapping criterion is easily satisfied (Line 14, Algorithm 3). Although the swapping criterion is satisfied and a pattern $g$ is hence swapped in, the pattern quality may be very low so that $g$ will be swapped out finally. Obviously, frequently swapping patterns does harm to algorithm performance. To address this problem, we develop *Initial Pattern Selection* (IPS) technique motivated by the following observation.

OBSERVATION II. *The initial k patterns generated by traversing the search space in DFS manner share common substructure and lead to low loss score. While patterns generated with Breadth-First Search (i.e., BFS) on the search space tend to be structurally dissimlar to each other and may imply higher loss score.*

Specifically, IPS starts at the first node in first level (e.g., $g_{1,1}$, Figure 5) of the search space. Then, it explores descendant nodes in second level (e.g., $g_{2,1}$) if the extended graph (e.g., $g_{2,1}$) has higher benefit score. The process repeats until no higher benefit score is obtained or the desirable number of edges (i.e., $E_{max}$) is derived. As a result, the first pattern is obtained. After that, IPS adopts the same technique to generate patterns rooted at the other nodes in first level (e.g., $g_{1,2}$ and $g_{1,3}$, Figure 5). Once all patterns rooted at nodes in first level are generated, the top-$k$ patterns with maximum coverage are selected as initial pattern set.

The improved TED algorithm with IPS (denoted by TED_IPS) is developed. The only difference between TED_IPS and TED_BASE is that TED_IPS generates initial patterns $\mathcal{P}$ with IPS instead of initializing it with an empty set (Line 1, Algorithm 3).

## 5.3 Putting Things Together (TED)

The complete TED algorithm integrates both two optimizations. Specifically, IPS is first introduced to improve the initial loss score. Then, PRM takes a prudent strategy to enumerate potential promising subgraphs. Theorem 4 ensures that the approximation ratio of our TED algorithm is lower bounded by 1/4.

THEOREM 4. *Let $\mathcal{P}_{opt}$ be an optimal solution to top-k edge-diversified patterns discovery problem. The approximation ratio of patterns $\mathcal{P}$ generated by TED (and TED_BASE) is bounded by $\frac{|Cov(\mathcal{P},D)|}{|Cov(\mathcal{P}_{opt},D)|} \geq \frac{1}{4}$.*

PROOF. We prove it by reducing our problem from max k-cover problem [22], which has a $\frac{1}{4}$-approximation solution when the swapping strategy is adopted [22]. Observe that the problem has the same setting as our problem if all promising patterns are generated. In addition, the same swapping strategy [22] is adopted by default. Hence, the approximation ratio of patterns $\mathcal{P}$ generated by TED is bounded by $|Cov(\mathcal{P},D)| / |Cov(\mathcal{P}_{opt},D)| \geq \frac{1}{4}$. □

THEOREM 5. *Worst case time and space complexities of TED are $O(|D|2^{max(V(G))^2})$ and $O(max(E(G))|D|)$, where $max(V(G))$ (resp. $max(E(G))$) is maximum number of vertices (resp. edges) in graph $G \in D$, and $|D|$ is number of graphs in D.*

PROOF. Compared to TED_BASE, TED introduces two optimizations. Since time complexity of *initial pattern selection* is not larger than that of subgraph enumeration and its space complexity is $O(kE_{max})$, worst case time and space complexities of TED will not increase after applying *initial pattern selection*. The same conclusion is made when applying *promising right-most extension*. Overall, worst case time and space complexities of TED are $O(|D|2^{max(V(G))^2})$ and $O(max(E(G))|D|)$, respectively. □

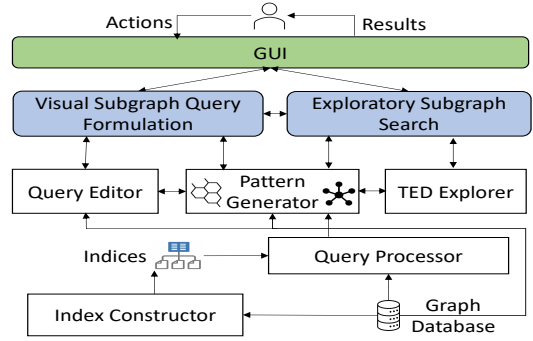

Figure 7: The architecture of VINCENT.

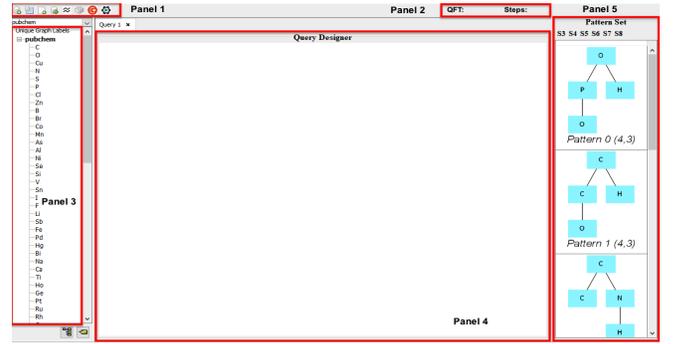

Figure 8: The interface of VINCENT.

## 6 THE APPLICATION POTENTIALS

In this section, we employ the demonstration system VINCENT [17] to illustrate the application potentials of edge-diversified patterns in visual subgraph query formulation and exploratory subgraph search. Figure 7 shows the architecture of VINCENT, which consists of five modules, Index Constructor, Query Processor, Pattern Generator, Query Editor, and TED Explorer. In particular, Pattern Generator takes as input a graph database $D$ and parameters (e.g., $k$), and generates top-$k$ edge-diversified patterns $\mathcal{P}_D$, which are displayed on the GUI (e.g., Panel 5, Figure 8). Query Editor provides a canvas (e.g., Panel 4, Figure 8) to users to visually formulate a query $Q$ by dragging-and-dropping patterns $\mathcal{P}_D$ and node labels in $D$ (see Section 6.1). Once the query $Q$ is formulated, Query Processor can efficiently process the query with the help of Index Constructor to generate subgraph matching results $R$. In particular, Query Processor performs subgraph isomorphism checking after Index Constructor filters out irrelevant results with $A^2F$ and $A^2I$ indices [47]. TED Explorer allows users to customize some parameters such as $k$, invokes Pattern Generator to generate edge-diversified patterns $\mathcal{P}_R$ for the query results $R$ based on these parameters, and finally enables users to explore the summarized query results (i.e., $\mathcal{P}_R$) (see Section 6.2). Based on these modules, the VINCENT system enables users to interactively experience visual subgraph query formulation and exploratory subgraph search, both of which are elaborated as below, with a focus on how edge-diversified patterns play a vital role in generating summary for the underlying database $D$ (i.e., patterns $\mathcal{P}_D$) and query results $R$ (i.e., patterns $\mathcal{P}_R$), respectively.

## 6.1 Visual Subgraph Query Formulation

Given a query $Q$ and a set of data graphs $D$, subgraph search/query is to retrieve the data graphs $R = \{G_i\}$ where $G_i \in D$ contains $Q$ (*i.e.*, $Q$ is subgraph isomorphic to $G_i$). It consists of two steps. The first step is subgraph query formulation (*i.e.*, how to construct the query graph $Q$). The second step is subgraph query processing (*i.e.*, how to find these matched graphs $R$, as discussed above).

In contrast to declarative query languages (*e.g.*, SPARQL and Cypher), visual subgraph query formulation helps non-programmers to take advantage of graph querying frameworks through a visual query interface (*also known as* GUI) for query construction. The core component of this interface is a set of subgraph patterns that allow users to construct multiple nodes and edges in a query $Q$ by performing a single click-and-drag action (*i.e.*, *pattern-at-a-time mode*) instead of iterative construction of edges one-at-a-time (*i.e.*, *edge-at-a-time mode*). Users are also allowed to delete nodes and edges from a partially constructed query. The *pattern-at-a-time mode* can greatly decrease the time taken to visually construct $Q$. In VINCENT, as shown in Figure 8, edge-diversified patterns are displayed in Panel 5, and Panel 3 displays the nodes of the underlying database. Users can visually formulate a query by dragging and dropping patterns from Panel 5 and nodes from Panel 3 to Panel 4.

## 6.2 Exploratory Subgraph Search

Subgraph search focuses on "lookup" retrieval with the assumption that users have a clear query intent (*i.e.*, know the exact structure of query $Q$) and sufficient knowledge of the underlying database $D$ to accurately construct $Q$. However, this assumption may become impractical as the graph database evolves. Exploratory subgraph search [19, 20, 51] alleviates this problem by supporting not only lookup retrieval but also exploratory search, which allows users to *iteratively or progressively formulate queries* and *explore the query results*. In this process, an end user becomes more familiar with the content and finally identifies the exact query $Q$. Suppose an user wants to query $Q$, she may not have the complete query structure "in her mind" at the very beginning. As VINCENT displays edge-diversified patterns $\mathcal{P}_D$ on the GUI, she may find a pattern $p$ interesting while browsing the pattern set, and initiate her query with $p$ (*i.e.*, $Q_0 = p$). By observing the query results of $Q_0$, she may iteratively construct $Q_1, Q_2, ...$ and finally $Q$. Without the help of these patterns, such a bottom-up search is not likely to happen.

Edge-diversified patterns can help in this process because the query results $R$ may contain a huge number of graphs, which hinders gaining insights from the results. To address this, VINCENT presents TED Explorer. In particular, for user-specified parameters such as $k$ and maximum (resp. minimum) pattern size MaxE (resp. MinE), TED Explorer invokes Pattern Generator to generate edge-diversified patterns $\mathcal{P}_R$ for query results $R$, which can be cast as a summary of query results and displayed on the interface for a better exploration experience.

## 7 PERFORMANCE STUDY

In this section, we investigate the performance of TED and report the key findings. TED is implemented with Java (JDK1.8). All experiments are conducted on a 64-bit Windows desktop with AMD Ryzen 5 3500X 6-Core CPU (3.6GHz) and 32GB of main memory.

Table 2: Datasets.

| Datasets | $E_{max}$ | $V_{max}$ | $E_{avg}$ | $V_{avg}$ | $|D|$ |
|---|---|---|---|---|---|
| AIDS | 251 | 222 | 27.3 | 25.4 | 40K |
| eMol | 104 | 100 | 15.9 | 15.5 | 10K |
| PubChem | 838 | 801 | 43.8 | 42.3 | 1M |

### 7.1 Experimental Setup

**Datasets.** The experiments are conducted on three datasets. (1) The dataset AIDS antiviral [6] consists of 40,000 (40K) data graphs. We also use AIDSL to denote the labeled AIDS where each graph is with labeled edges (*i.e.*, bonds). (2) The dataset *PubChem* [7] consists of many chemical compound graphs. Unless otherwise stated, *PubChem* refers to the 23K dataset. Other variants used are 100K, 300K, 500K and 1 million (1M). (3) The dataset *eMol* [8] consists of 10K chemical compounds. Note that <Y><X> are used to denote variants of various datasets, where $Y$ and $X$ refer to the name of the dataset and the number of graphs used, respectively. For example, AIDS10K refers to the dataset AIDS consisting of 10K data graphs. Their statistics are given in Table 2 where $E_{max}$ (resp. $V_{max}$), $E_{avg}$ (resp. $V_{avg}$) and $|D|$ indicate maximum number of edges (resp. vertices), average number of edges (resp. vertices), and number of graphs, respectively.

**Baselines.** We compare TED against the proposed baselines and their variants: (1) enumerating all subgraphs and then performing greedy search (ALL$_g$, Algorithm 1), (2) integrating enumeration of all subgraphs and swapping-based search (ALL$_t$, *i.e.*, a variant of ALL$_g$), (3) enumerating all frequent subgraphs and then performing greedy search (FSG$_g$, Algorithm 2), and (4) integrating enumeration of all frequent subgraphs and swapping-based search (FSG$_t$, *i.e.*, a variant of FSG$_g$). We also compare TED with its variants (BASE and PRM) where (1) BASE: TED_BASE (Algorithm 3), (2) PRM: BASE + TED_PRM and (3) TED: PRM + TED_IPS. To illustrate the application potentials of edge-diversified patterns, we further compare TED against (1) CATAPULT [16], the state-of-the-art visual query formulation method, and (2) top-$k$ frequent patterns (denoted as FS).

**Parameter settings.** Unless specified otherwise, we set $k = 5$ and $E_{max} = 10$.

**Performance measures.** We use the following measures for performance evaluation: (1) *Processing Time (in second):* time taken to generate the patterns. Note that INF is used to denote the case that a test does not stop in a time limit (10000 seconds) or memory limit. In the experiment, we use log scale for the time. (2) *Coverage Rate:* the coverage rate of patterns to the total number of edges in a database $D$; (3) *Index Time (in second):* time taken to maintain PES-index; (4) *Index Size (in kilobyte):* space consumption for storing the PES-index. (4) *Query Formulation Time* (QFT): time taken to formulate a query; (5) *Steps*: steps taken to formulate a query; (6) *Reduction Ratio* (RR): Ratio of reduced steps when patterns derived from TED are used, RR = $\frac{Steps_X - Steps_{TED}}{step_X}$ where $step_X$ and $step_{TED}$ are the *minimum* number of steps required to construct a query $Q$ when patterns $\mathcal{P}$ derived from the approach $X$ and TED are used, respectively. As we discussed above, $X$ could be CATAPULT or top-$k$ frequent patterns mining. RR > 0 implies that $\mathcal{P}$ derived from TED required fewer steps than $X$. Note that we follow the same assumptions in [16] to estimate $Steps_X$ and $Steps_{\text{TED}}$: (1) a pattern $p \in \mathcal{P}$ can be used to construct the query $Q$ iff $p \subseteq Q$; (2) when multiple patterns are used to construct $Q$, their corresponding isomorphic subgraphs in $Q$ do not overlap. We shall remove these assumptions in the user study, where users are allowed to modify (e.g., delete nodes/edges) patterns when they are used for query formulations.

---
[6] https://wiki.nci.nih.gov/display/NCIDTPdata/AIDS+Antiviral+Screen+Data
[7] ftp://ftp.ncbi.nlm.nih.gov/pubchem/Compound/CURRENT-Full/SDF/
[8] https://www.emolecules.com/info/plus/download-database

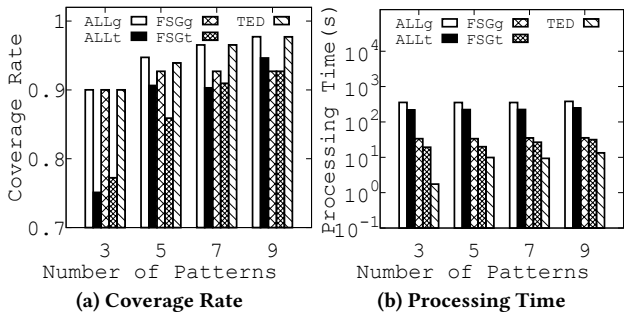

(a) Coverage Rate   (b) Processing Time

Figure 9: Effect of Number of Patterns.

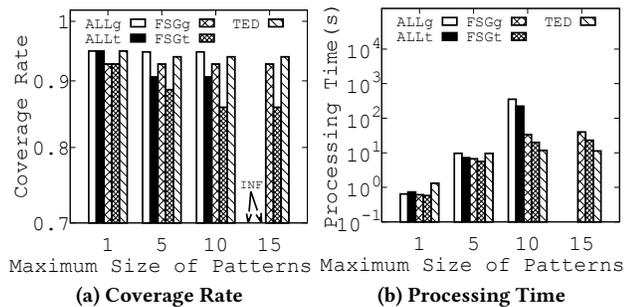

(a) Coverage Rate   (b) Processing Time

Figure 10: Effect of Maximum Size of Patterns.

### 7.2 Experimental Results

#### 7.2.1 Comparison with Baselines.

**Question 1:** *How does TED perform compared to the baselines in the default parameter settings? And do different parameters (i.e., $E_{max}$ and $k$) affect the results?*

**Result 1.** TED outperforms baselines in terms of both Processing Time and Coverage Rate in the default parameter settings (see Exp 2). In general, TED is comparable to $ALL_g$ and outperforms other methods in terms of coverage rate, and requires less processing time (see Exp 1).

**Exp 1: Setting of Maximum Size and Number of Patterns.** To evaluate the performance of TED, we first perform an evaluation to determine parameter settings. We vary the number of patterns (i.e., $k$) on AIDS5K. Figure 9 plots the results. In general, the coverage rate and processing time increase with $k$, since more patterns are introduced and higher coverage will be obtained as $k$ increases. The methods based on greedy search (i.e., $ALL_g$ and $FSG_g$) generally show better coverage rate and more processing time compared to swapping-based search methods (i.e., $ALL_t$ and $FSG_t$), as the former needs to store all (resp. frequent) subgraphs for further searching. However, TED is always comparable to $ALL_g$ and better than other methods in terms of coverage rate (Figure 9(a)), although it is based on a swapping-based search. As shown in Figure 9(b), TED requires less processing time than other methods on all settings. In the following experiments, we set $k = 5$.

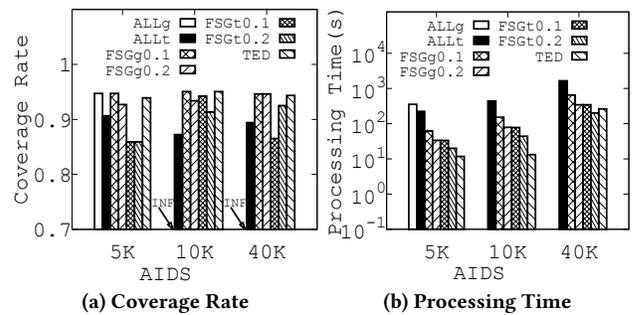

(a) Coverage Rate   (b) Processing Time

Figure 11: Baseline comparison on AIDS.

We also study the performance of TED with different maximum sizes of patterns ($E_{max}$) on AIDS5K. As shown in Figure 10, TED is comparable to $ALL_g$ and better than other methods in terms of coverage rate, but requires less processing time in most cases, especially for larger $E_{max}$. Intuitively, $ALL_g$ can obtain a better coverage rate compared to other methods except TED since all subgraphs are enumerated and stored, which makes it memory- and time-consuming (e.g., INF for $E_{max} = 15$). In addition, as $E_{max}$ increases, the search space is enlarged, and hence the processing time of these methods increases. Note that the coverage rate fluctuates within a narrow range, which is due to the label distribution of edges in the database. We set $E_{max} = 10$ in the following experiments.

**Exp 2: Comparison with Baselines.** Next, we compare TED with baselines and their variants, i.e., $ALL_g$, $ALL_t$, $FSG_g$, and $FSG_t$ in the default setting, and report the results on AIDS dataset in Figure 11. The results of FSG-based algorithms (i.e., $FSG_g$ and $FSG_t$) with various supports (0.1 and 0.2) are also reported. As depicted in Figure 11(a), in terms of coverage rate, TED outperforms other baselines and is comparable to $ALL_g$, which incurs INF on AIDS10K and AIDS40K. As the size of dataset increases from 5K to 40K, processing time of $ALL_g$ increases dramatically, while that of our TED algorithm increases steadily to less than 4 minutes on AIDS40K, as shown in Figure 11(b). Hence, TED outperforms baselines and their variants.

To investigate the effect of maximum number of nodes in a graph, we compare TED with baselines on DS = $\{D_{(0,20]}, D_{(20,50]}, D_{(50,80]}, D_{(80,801]}\}$ [9] where $D_{(r,l]}$ ($|D_{(r,l]}| = 1000$) represents the graphs in *PubChem* whose node sizes are in the range $(r, l]$. Figure 12 depicts the results. In general, TED consistently shows comparable coverage rate to greedy-search based methods, whose processing time dramatically increases with the maximum number of nodes in a graph and even incurs INF. We can also observe that the coverage rate slightly increases with the maximum number of nodes. This is because total number of edges in each dataset increases more raipdly than the number of uncovered edges.

Finally, we compare them with the optimal solutions on small datasets (i.e., *PubChem*100 and *AIDS*100) and report the results in Figure 13. We observe that ratio of the coverage rate of TED to that of optimal solution (denoted by OPT in the figure) is no less than 0.945, which is far better than the theoretical approximation ratio.

---

[9]As shown in Table 2, maximum number of nodes in PubChem is 801.

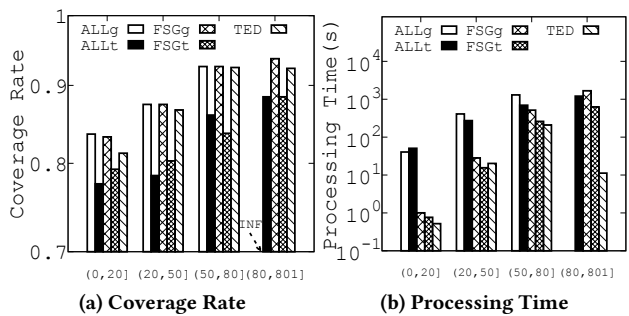

**(a) Coverage Rate**  **(b) Processing Time**

Figure 12: Effect of Maximum Number of Nodes in a Graph.

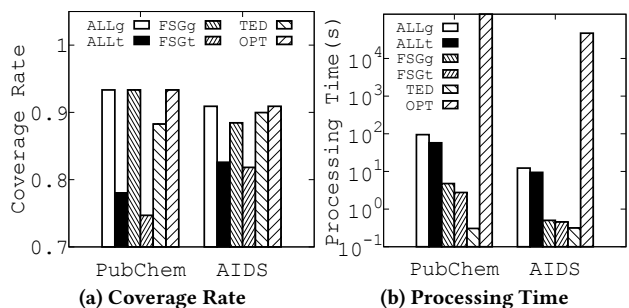

**(a) Coverage Rate**  **(b) Processing Time**
Figure 13: Baseline comparison on *PubChem*100 and *AIDS*100.

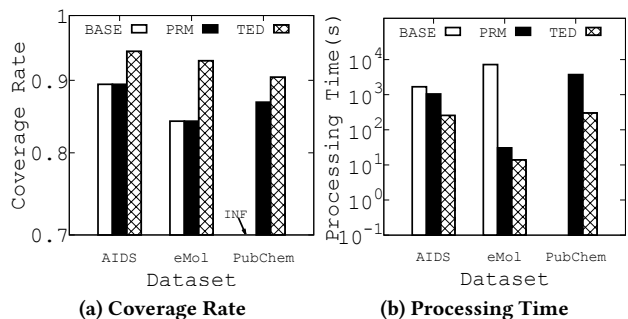

**(a) Coverage Rate**  **(b) Processing Time**
Figure 14: Effect of Optimization Strategies.

To sum up, TED outperforms baselines in terms of both coverage rate and processing time.

#### 7.2.2 Effectiveness and Scalability Evaluation.
**Question 2:** *Are the presented techniques (optimizations and swapping criteria) effective? Are the space and time taken to store/maintain PES-Index acceptable?*
**Result 2.** Both optimizations and swapping criteria are effective as the former can reduce processing time without decreasing coverage rate (see Exp 3), and the latter can facilitate TED in coverage rate and processing time, regardless of which swapping threshold is used (see Exp 5). In addition, the space and time taken to store/maintain PES-Index are acceptable (see Exp 4).

**Exp 3: Effect of Optimization Strategies.** We further evaluate the effectiveness of optimization strategies by comparing BASE and PRM with TED where all optimization strategies are used. The results are presented in Figure 14. Compared to BASE, and PRM, TED reduces processing time but does not decrease coverage rate, as it eliminates unpromising patterns without sacrificing promising ones. Thanks to all adopted optimization strategies, TED is the best one in terms of coverage rate and processing time compared to BASE and PRM. In addition, we can observe that the processing time of BASE, PRM, and TED shows a decreasing trend, which further justifies the effectiveness of optimization strategies.

**Table 3: Size of PES-Index**

| Dataset | AIDS | | eMol | | PubChem | |
|---|---|---|---|---|---|---|
| | 10K | 40K | 5K | 10K | 10K | 23K |
| Index Size(KB) | 234 | 1008 | 89 | 157 | 428 | 1157 |
| Index/Graphs (%) | 5.39 | 5.31 | 5.40 | 5.39 | 5.80 | 7.58 |

**Table 4: Maintenance Time of PES-Index**

| Dataset | AIDS | | eMol | | PubChem | |
|---|---|---|---|---|---|---|
| | 10K | 40K | 5K | 10K | 10K | 23K |
| Index Time(s) | 0.5 | 1.88 | 0.25 | 0.37 | 1.1 | 2.85 |
| Index Time/Total (%) | 6.86 | 1.00 | 4.12 | 3.63 | 0.78 | 1.39 |

**Exp 4: PES-Index Test.** In this experiment, we test the size and maintenance time of PES-Index. As presented in Table 3, PES-Index size increases with the size of the dataset. Note that in comparison with the size of dataset, PES-Index size is very small since only five components are stored in PES-Index. In particular, for the larger datasets (*e.g.*, PubChem23K), the space taken to store PES-Index is only 5.31% and 7.58% of the size of the underlying dataset.

We also report the maintenance time of PES-Index in Table 4. Observe that as the size of dataset increases, maintenance time increases accordingly but makes up less than 7% of the total processing time. For example, maintenance time of PES-Index on AIDS40K and PubChem23K are 1% and 1.39% of the corresponding total processing time, respectively.

**Exp 5: Effect of Swapping Criteria.** We investigate the effect of different swapping criteria (see Section 4), namely $Swap_1$, $Swap_2$, and $Swap_\alpha$. The results are reported in Figure 15. In general, TED outperforms the baselines in terms of both coverage rate and processing time, no matter what swapping criteria are used. Although $FSG_g$ with $Swap_2$ shows a higher coverage rate in eMol, the results generated by $FSG_g$ are data-dependent and are not theoretically guaranteed. Furthermore, it may incur INF in larger datasets (*e.g.*, PubChem). This experiment further justifies the effectiveness of TED algorithm.

#### 7.2.3 Application Potentials Evaluation.
**Question 3:** *Whether top-k edge-diversified patterns can facilitate existing or potential applications? These patterns may contain infrequent subgraphs, why do infrequent patterns remain useful?*
**Result 3.** Top-$k$ edge-diversified patterns can facilitate both visual subgraph query formulation and exploratory subgraph search (see Exp 6). While frequent subgraphs are often useful, infrequent patterns can also facilitate applications where queries are not necessarily frequent (see Exp 7).

**Exp 6: User Study.** In this section, we conduct user studies based on VINCENT to evaluate the application potentials of edge-diversified patterns. We recruit 15 unpaid volunteers (ages from 20 to 32) in accordance to HCI research that recommends at least 10 participants [27, 28]. Before conducting any user study, all volunteers, who have backgrounds in chemistry, chemical engineering, CS, biology, were trained to use VINCENT. More details on how to use VINCENT are provided in Section 6.

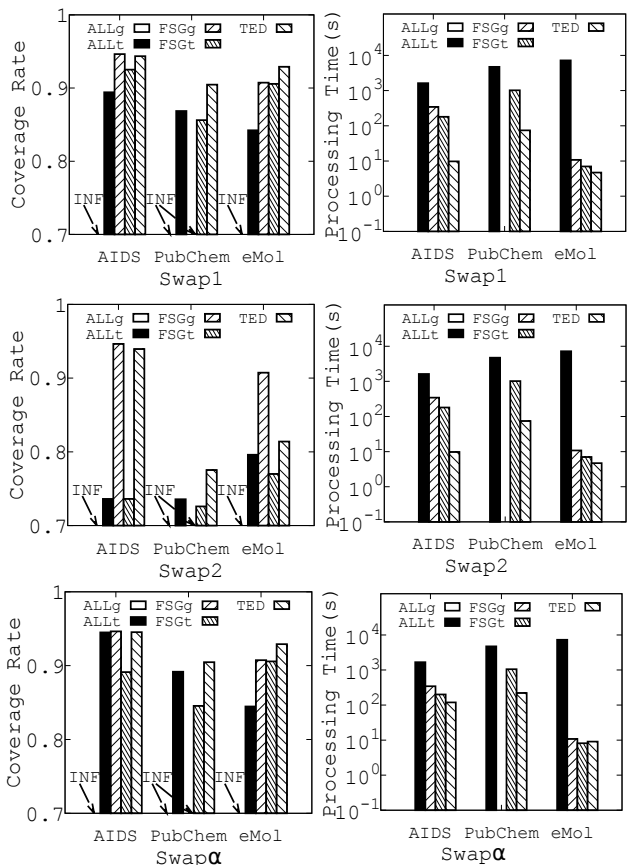

Figure 15: Effect of Swapping Criteria.

**Visual Query Formulation.** The first study aims to evaluate the application potential in visual query formulation (VQF). To this end, we compare edge-diversified patterns with canned patterns provided by CATAPULT [16] (the state-of-the-art VQF method), and frequent patterns (denoted as FS). In particular, we first follow the existing work [16] to select 5 queries (see Table 5) of size in the range [30-62], which span a variety of structures (cycles, carbon chains, etc.) and contain different vertex labels. Each query in a dataset (AIDS or *PubChem*) is associated with a unique identifier (*also known as* CID, Table 5) in the *PubChem* repository [10] provided by National Institutes of Health (NIH). Second, each pattern set (TED or CATAPULT or FS) is displayed on the GUI (Panel 5 in Figure 8) for VQF. Volunteers are allowed to visually formulate queries by dragging and dropping patterns from Panel 5 and nodes from Panel 3 to Panel 4 to formulate queries. *Query Formulation Time* (QFT) and *Steps* taken are recorded in Panel 2 (Figure 8).

Figure 16 reports the results on PubChem and AIDS. Observe that compared to CATAPULT and FS, TED facilitates more efficient query formulations (shorter QFT and fewer steps). In addition, we can also observe that some queries (*e.g.,* $Q_5$ in Figure 16(b)) compared to other queries (*e.g.,* $Q_1$ in Figure 16(b)) benefit more from TED, *i.e.,* TED can save more QFT and *Steps* taken on $Q_5$ than those on $Q_1$ when compared to FS and CATAPULT. The main reason lies in that TED enjoys more patterns used for formulating $Q_5$ ($|\mathcal{P}_U|$(FS)=2 vs $|\mathcal{P}_U|$(CATAPULT)=3 vs $|\mathcal{P}_U|$(TED)=6) than $Q_1$ ($|\mathcal{P}_U|$(FS)=1 vs $|\mathcal{P}_U|$(CATAPULT)=2 vs $|\mathcal{P}_U|$(TED)=3), as shown in Table 6. Note that the number of patterns used to formulate a query $Q$ indicates how many patterns can be used to cover different parts of $Q$ so that it can enjoy the *pattern-at-a-time* mode (see Section 6.1) and thus reduce time/step to visually construct $Q$.

[10]https://pubchem.ncbi.nlm.nih.gov/

**Table 5: Queries. CID is the unique identifier in the PubChem repository provided by National Institutes of Health (NIH).**

| Queries | CID, Pubchem ($|E|$) | CID, AIDS ($|E|$) |
|---|---|---|
| $Q_1$ | 169132 ($|E| = 34$) | 135398740 ($|E| = 32$) |
| $Q_2$ | 20497364 ($|E| = 30$) | 565070 ($|E| = 34$) |
| $Q_3$ | 493570 ($|E| = 47$) | 102034018 ($|E| = 35$) |
| $Q_4$ | 135398658 ($|E| = 52$) | 14852846 ($|E| = 30$) |
| $Q_5$ | 3324 ($|E| = 42$) | 154402349 ($|E| = 62$) |

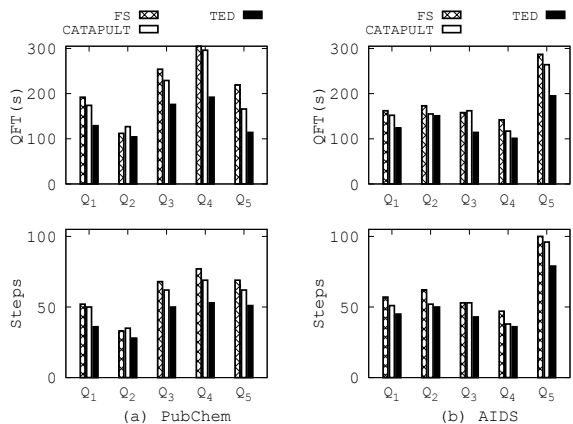

Figure 16: Query Formulation Time (QFT) and Steps.

**Table 6: Number of Patterns Used in VQF ($|\mathcal{P}_U|$). "Yes" indicates that at least one infrequent pattern can be used.**

| | PubChem | | | AIDS | | |
|---|---|---|---|---|---|---|
| Queries | FS | CATAPULT | TED | FS | CATAPULT | TED |
| $Q_1$ | 2 | 2 | **5** | 1 | 2 | **3** |
| $Q_2$ | 3 | 3 | **5 (Yes)** | 1 | 1 | **2** |
| $Q_3$ | 3 | 4 | **6 (Yes)** | 2 | 1 | **4** |
| $Q_4$ | 4 | 5 | **7 (Yes)** | 1 | 2 | **3** |
| $Q_5$ | 2 | 2 | **5 (Yes)** | 2 | 3 | **6 (Yes)** |

We use "Yes" in Table 6 to denote that at least one infrequent ($sup_{min} < 0.2$) edge-diversified pattern can be used in VQF and find that infrequent patterns can also facilitate VQF. This justifies the rationality of top-$k$ edge-diversified patterns discovery problem.

**Exploratory Subgraph Search.** We also conduct a user study on exploratory subgraph search. As exploratory search activities have no predetermined goals and are considered as open-ended [20], in this experiment, queries are user-specified rather than predetermined. As discussed in Section 6.2, volunteers are allowed to enjoy not only the bottom-up search but also TED Explorer, which helps to explore the query results. Therefore, we divide volunteers into two groups to iteratively construct user-specified queries and explore the query results. The first group uses VINCENT with TED Explorer to display patterns (*e.g.,* edge-diversified patterns), while the second group cannot use TED Explorer. Compared to the second group, we observe that the first group takes 20% less time when displaying edge-diversified patterns, and 10%–14% less time when displaying FS and CATAPULT. This means edge-diversified patterns enhance not only exploration efficiency but query experience.

Table 7: Patterns with Biological Importance. CID is the unique identifier in the PubChem repository provided by National Institutes of Health (NIH).

| Pattern Set | Patterns with Biological Importance | Total |
| --- | --- | --- |
| FS | X-Methylpentane (*e.g.*, CID 7892)), Carbon Chains (*e.g.*, CID 7843), CID 6556, CID 241, CID 6360 | 5 |
| CATAPULT | Carbon Chains, CID 7282, CID 16665, CID 119440, CID 6380, CID 19660, CID 702, CID 11507 | 8 |
| TED | Carbon Chains, CID 10903, CID 12230, CID 7964, CID 11473, CID 3034819, CID 702, CID 12338 | 8 |

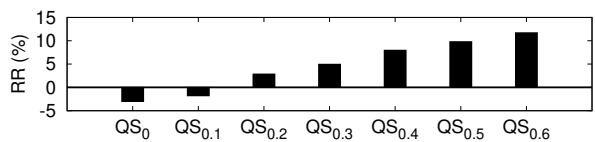

Figure 17: Reduction Ratio (RR).

**Other findings.** The most appealing thing to us is that top-$k$ edge-diversified patterns generated by TED contain not only patterns with statistical significance (*e.g.*, frequent patterns) but also patterns with biological importance. In this paper, a pattern is said to have biological importance if it exists in the *PubChem* repository [11], which is maintained by the National Institutes of Health (NIH) and contains millions of chemical molecules and their activities against biological assays. Table 7 reports these patterns' unique identifies (*i.e.,* CID). We can observe that compared to FS, TED and CATAPULT contain more such patterns (8 vs 5). For example, the CID 11473 in TED is an important organic compound *Nitrosobenzene*, which is one of the prototypical organic nitroso compounds. Therefore, TED may open up new opportunities in bioinformatics, drug design, etc.

**Exp 7: Effect of Queries.** As discussed above, infrequent edge-diversified patterns can also facilitate visual query formulation (VQF). The problem becomes *why infrequent patterns remain useful?* We answer this question by investigating the *ratio of reduced steps* for VQF (*i.e.,* RR = $\frac{Steps_{FS} - Steps_{TED}}{step_{FS}}$, see Section 7.1) between TED and top-$k$ frequent patterns (FS) on the query set $QS_\rho$ ($|QS_\rho| = 100$), where $\rho$ is the fraction of queries that are infrequent. When $\rho = 0$, all queries in $QS_\rho$ are frequent. We vary $\rho$ in $\{0, 0.1, 0.2, ..., 0.6\}$ and report the results in Figure 17. Obviously, TED underperforms FS on $QS_0$ (RR < 0) as TED contains a mixture of frequent and infrequent patterns. Nevertheless, RR increases with $\rho$ and is larger than 0 at $\rho = 0.2$. This indicates that in terms of facilitating VQF of infrequent queries, the performance of TED improves as the ratio of infrequent queries increases, which explains the reason why infrequent patterns remain useful.

## 8 RELATED WORK

Subgraph enumeration listing all subgraphs or counting all instances of a particular graph in a graph database has been extensively studied in the literature [1, 2, 4, 5, 29, 32–39]. Instead of enumerating all subgraphs, frequent subgraph mining (FSM) is to generate only frequent subgraphs. Existing FSM methods [6, 7, 9–12] consider two settings, transactional and single graph such as [12]. Rather than enumerating all frequent subgraphs, some work generate only representative subgraphs [16, 30, 31, 40, 40–43, 49], none of them focus on top-$k$ edge-diversified problems and their solutions can not be directly adapted for it. Subgraph matching [36, 38, 48] finds the matched subgraphs in a data graph for a given query graph, which is a pattern matching problem instead of pattern mining problem. Diversity problem is the most germane to this research, which seeks for search results with diversity[13, 44–46, 50, 52]. [44] presents diversity-aware search method of relevant documents. Fan et al. [13] aims to retrieve diversified top-$k$ matches for a given vertex. The problem of finding redundancy-aware maximal cliques is considered by [45]. In addition, [46] further studies the diversified top-$k$ clique search problem, which is to find $k$ maximal cliques in a data graph so that the maximum number of vertices are included. Given a query graph, [52] retrieves diversified top-$k$ matches to cover more vertices. To the best of our knowledge, our work is the first one to study top-$k$ edge-diversified patterns discovery problem.

Top-$k$ edge-diversified patterns discovery problem is also related to the set cover problem [22]. Given a universal set $\mathcal{U}$ of $n$ elements and a collection $\mathcal{S} = \{S_1, S_2, ..., S_m\}$ of $m$ subsets of $\mathcal{U}$ ($\bigcup_i S_i = \mathcal{U}$), the set cover problem is to find as few subsets as possible from $\mathcal{S}$ such that their union covers $\mathcal{U}$. The edge cover problem is a special case of set cover problem, in which the elements of the universe are vertices and each subset covers exactly two elements. As we discussed in Section 2.2, they are not the same as the edge-diversified patterns discovery problem.

## 9 CONCLUSION

In this work, we investigate the top-$k$ edge-diversified patterns discovery problem, which is to find $k$ subgraphs from a graph database such that the maximum number of edges in the database are covered. Maximizing total covered edges requires that patterns should have not only high subgraph coverage (*i.e.,* more data graphs are covered by the patterns) but also high diversity (*i.e.,* patterns are diverse to each other to cover different parts of data graphs). This problem may nurture a lot of applications such as visual query formulation. To address this problem, we present a novel framework called TED and an efficient index structure. In addition, three optimization strategies are developed to improve the performance. To handle even larger graph databases, a lightweight version called TEDLITE is further designed. TED requires limited memory and achieves a guaranteed approximation ratio. Extensive experimental results justify the superiority of TED over baseline solutions.

## 10 ACKNOWLEDGMENT

This work was supported by National Natural Science Foundation of China (Grant No: 62072390, 62102334), and the Research Grants Council, Hong Kong SAR, China (Grant No: 16202722, 15222118, 15218919, 15203120, 15226221, 15225921, and C2004-21GF). The research work described in this paper was partially conducted in the JC STEM Lab of Data Science Foundations funded by The Hong Kong Jockey Club Charities Trust.

[11]https://pubchem.ncbi.nlm.nih.gov/